\documentclass[useAMS,usenatbib,a4paper]{mn2e}

\usepackage[left=2cm,top=2cm,right=2cm,bottom=2cm]{geometry}
\usepackage{amssymb,epsfig,color}
\usepackage{amsmath}
\usepackage{graphicx}
\usepackage{multirow}
\usepackage{xspace}
\usepackage{hyperref}
\usepackage{textcomp}
\usepackage{caption,subcaption}

\title[RoboPol Pipeline and Control System]{The RoboPol Pipeline and Control
System}
\author[O.\,G.~King et al.]{O.\,G.~King$^{1}$\thanks{E-mail:ogk@astro.caltech.edu}, 
D.~Blinov$^{2,7}$, 
A.\,N.~Ramaprakash$^{3}$,
I.~Myserlis$^{4}$,
E.~Angelakis$^{4}$,
\newauthor
M.~Balokovi\'{c}$^{1}$, 
R.~Feiler$^{5}$, 
L.~Fuhrmann$^{4}$, 
T.~Hovatta$^{1,8}$, 
P.~Khodade$^{3}$, 
\newauthor
A.~Kougentakis$^{6}$,
N.~Kylafis$^{2,6}$, 
A.~Kus$^{5}$, 
D.~Modi$^{3}$,
E.~Paleologou$^{2}$,
\newauthor
G.~Panopoulou$^{2}$, 
I.~Papadakis$^{2,6}$,
I.~Papamastorakis$^{2,6}$,
G.~Paterakis$^{2}$, 
\newauthor 
V.~Pavlidou$^{6,2}$,
B.~Pazderska$^{5}$,
E.~Pazderski$^{5}$, 
T.\,J.~Pearson$^{1}$, 
C.~Rajarshi$^{3}$,
\newauthor
A.\,C.\,S.~Readhead$^{1}$, 
P.~Reig$^{6,2}$,
A.~Steiakaki$^{2}$,
K.~Tassis$^{2,6}$,
J.\,A.~Zensus$^{4}$
\\
$^{1}$Cahill Center for Astronomy and Astrophysics, California Institute of Technology, 1200 E
California Blvd, MC 249-17, \\Pasadena CA, 91125, USA\\
$^{2}$Department of Physics and Institute of Theoretical \& Computational Physics, University of
Crete, PO Box 2208, \\GR-710 03, Heraklion, Crete, Greece\\
$^{3}$Inter-University Centre for Astronomy and Astrophysics, Post Bag
4, Ganeshkhind, Pune - 411 007, India\\
$^{4}$Max-Planck-Institut f\"{u}r Radioastronomie, Auf dem H\"{u}gel
69, 53121 Bonn, Germany\\
$^{5}$Toru\'{n} Centre for Astronomy, Nicolaus Copernicus University, Faculty of
Physics, Astronomy and Informatics, \\Grudziadzka 5, 87-100 Toru\'{n}, Poland \\
$^{6}$Foundation for Research and Technology - Hellas, IESL, Voutes, 7110 Heraklion, Greece \\
$^{7}$Astronomical Institute, St. Petersburg State University,Universitetsky pr. 28,
Petrodvoretz, 198504 St. Petersburg, Russia \\
$^{8}$Aalto University Mets\"ahovi Radio Observatory, Mets\"ahovintie
114, 02540 Kylm\"al\"a, Finland
}

\begin{document}

\date{Accepted XXX. Received YYY; in original form ZZZ}

\pagerange{\pageref{firstpage}--\pageref{lastpage}} \pubyear{2013}

\maketitle

\label{firstpage}

\begin{abstract}
We describe the data reduction pipeline and control system for the RoboPol
project. The RoboPol project is monitoring the optical $R$-band magnitude and linear polarization
of a large sample of active galactic nuclei that is dominated by blazars. The pipeline calibrates
and reduces each exposure frame, producing a measurement of the magnitude and linear polarization of
every source in the $13'\times 13'$ field of view. The control system combines a dynamic scheduler,
real-time data reduction, and telescope automation to allow high-efficiency unassisted observations.
\end{abstract}


\section{Introduction} \label{sec:introduction}

The RoboPol project\footnote{\url{http://www.robopol.org/}} is monitoring the $R$-band optical
linear polarization and magnitude of a large sample of active galactic nuclei (AGN). The
statistically well-defined sample is
drawn from gamma-ray loud AGN detected by \emph{Fermi}
\citep{2010ApJS..188..405A,2012ApJS..199...31N} and is dominated by blazars, as described in
V. Pavlidou et al., in prep. The main science goal of the RoboPol project is to understand the link
in AGN between optical polarization behavior, particularly that of the electric vector position
angle (EVPA) (e.g., \citealt{Marscher:2008p2271,Abdo:2010p5253}), and flares in gamma-ray
emission.

The RoboPol polarimeter (A.N. Ramaprakesh et al. in prep.) is an imaging photopolarimeter that
measures the linear polarization and magnitude of all sources in the $13'\times13'$
field of view. It is installed on the $1.3$-m telescope at the Skinakas
Observatory\footnote{\url{http://skinakas.physics.uoc.gr/}} in Crete,
Greece. The large amount of observing time (four nights a week on average over the Skinakas 9-month
observing season) and long duration of the
project (at least three years) will generate a large amount of data, requiring a fully automated
data reduction pipeline and observing procedure. While the RoboPol instrument is optimized for
operation in the $R$-band, it can also observe in the $I$ and $V$-bands. All observations 
described in this paper were made with a Johnson-Cousins $R$-band filter.

Blazar emission at optical wavelengths is highly variable and the optical polarization events we aim
to characterize can occur very rapidly. This requires a flexible observing scheme capable of
responding to changes in the optical polarization of a source without human intervention.
In this paper we describe the data reduction pipeline and control system developed to meet these
requirements. It is organized as follows. The telescope and instrument are described in
Section~\ref{sec:TI}. The data reduction pipeline and its performance are described in
Section~\ref{sec:PL}, and the control system is described in Section~\ref{sec:CS}. We conclude in
Section~\ref{sec:conclusion}.

\section{Telescope and Instrument} \label{sec:TI}

\subsection{Telescope} \label{sec:TI:telescope}

The $1.3$-m telescope at the Skinakas Observatory ($1750\,$m, $23^{\circ}53'57''$E,
$35^{\circ}12'43''$N) has a modified Ritchey-Chr\'{e}tien optical system ($129\,$cm primary,
$45\,$cm secondary, $f/7.54$). It has an equatorial mount, built by DFM
Engineering\footnote{\url{http://www.dfmengineering.com/}}, with an off-axis guiding system. The
telescope is equipped with several other instruments in addition to the RoboPol polarimeter,
including an imaging camera, IR camera, and spectrograph.

Control of the telescope and its subsystems is spread over several computers. The guiding camera,
its focus control, the RoboPol filter wheel, and the RoboPol CCD camera are connected directly to
the main control computer. The secondary mirror focus control, the dome control, and the equatorial
mount control are connected to the telescope control system (TCS) computer, which interfaces with
the main control computer through a serial link. A third computer monitors the weather station.

\subsection{Instrument} \label{sec:TI:instrument}

\begin{figure}
 \centering
 \includegraphics[width=0.47\textwidth]{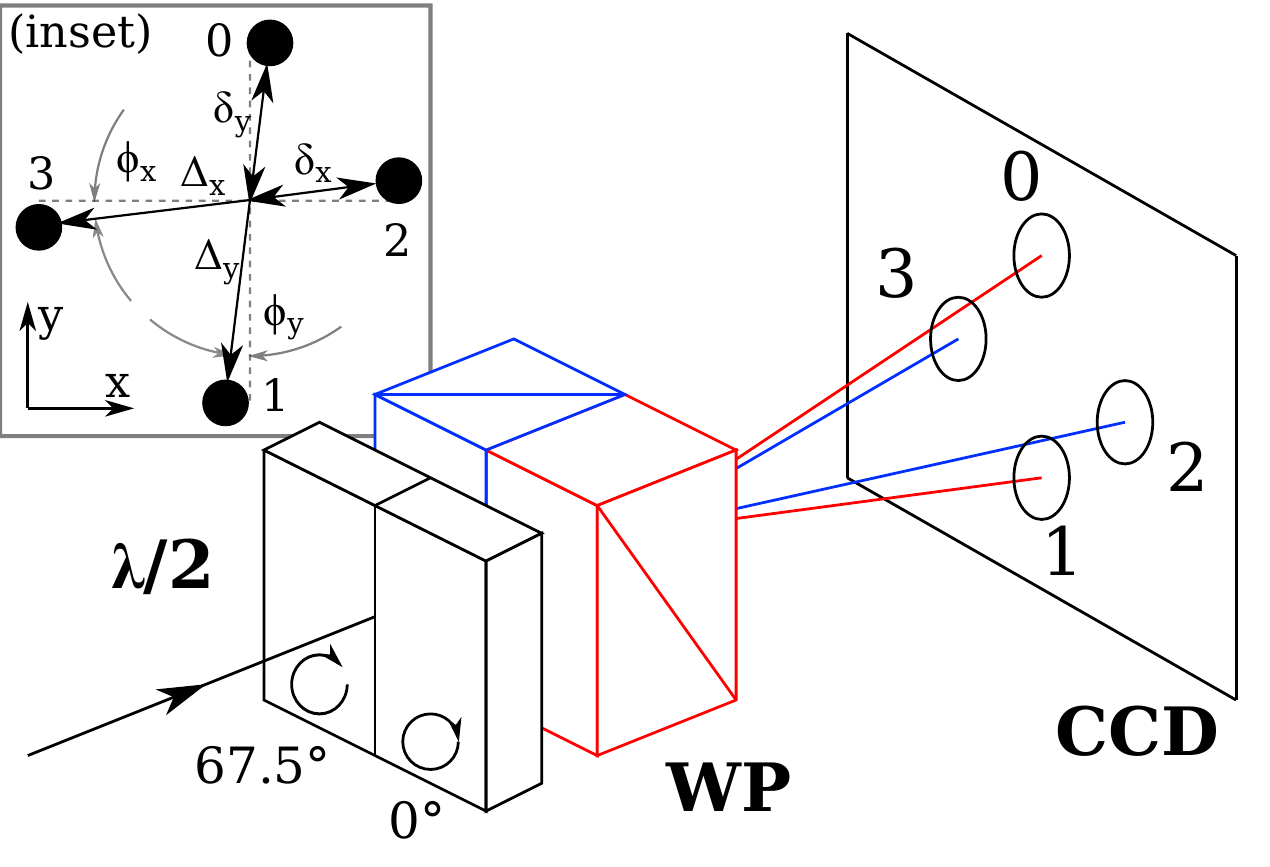}
 % robopol_drawing.pdf: 366x250 pixel, 72dpi, 12.91x8.82 cm, bb=0 0 366 250
\caption{Diagram showing the basic operation of the RoboPol instrument. The pupil of the instrument
is split in two, each half incident on a half-wave retarder followed by a Wollaston prism, labelled
$\lambda/2$ and WP respectively, with differing fast axis and prism orientations as indicated. The
blue pair split the rays horizontally to produce the spots labelled 2 and 3, while the red pair
produce the vertical spots 0 and 1. The linear polarization parameters are then calculated using
Eqn.~\ref{eqn:equation_for_stokes_q_u}. \ \ (inset) The pattern of spots at each position on the CCD
is described by this model. The distance between the spots $\Delta_x$ and $\Delta_y$, their distance
from the intersection point $\delta_x$ and $\delta_y$, and their angle with respect to the CCD axes
$\phi_x$ and $\phi_y$ all vary independently across the field. }
 \label{fig:optics_diagram}
\end{figure}

The RoboPol instrument (A.N. Ramaprakesh et al., in prep) is a 4-channel imaging
photopolarimeter designed with high observing efficiency and automated operation as prime goals. It
has no moving parts other than a filter wheel. Instead, as shown in Fig.~\ref{fig:optics_diagram},
the instrument splits the pupil in two -- each half incident on a half-wave retarder
followed by a Wollaston prism (WP). One prism is oriented such that it splits the rays falling on it
in the horizontal plane (blue prism and rays in Fig.~\ref{fig:optics_diagram}), while the other
prism's orientation splits them in the vertical plane (red in Fig.~\ref{fig:optics_diagram}).

Every point in the sky is thereby projected to four points on the CCD. The fast axis of the
half-wave
retarder in front of the first prism is rotated by $67.5^{\circ}$ with respect to the other
retarder. In the instrument reference frame the horizontal channel measures the $u=U/I$ fractional
Stokes parameter, while the vertical channel measures the $q=Q/I$ fractional Stokes parameter,
simultaneously, with a single exposure. This design
eliminates the need for multiple exposures with different half-wave plate positions,
thereby avoiding systematic and random errors due to sky changes between
measurements and imperfect alignment of rotating optical elements.

The expressions for the relative Stokes parameters and their
uncertainties are (see A.N. Ramaprakash et al. (in prep.)  for the derivation):
\begin{align}
\nonumber q = & \frac{N_{1}-N_{0}}{N_{0}+N_{1}} ,~~~
\sigma_q =  \sqrt{\frac{4(N_{1}^2\sigma_{0}^2+N_{0}^2\sigma_{1}^2)}{(N_{0}+N_{1})^4}}, \\
u =  & \frac{N_{2}-N_{3}}{N_{2}+N_{3}},~~~
\sigma_u = \sqrt{\frac{4(N_{3}^2\sigma_{2}^2+N_{2}^2\sigma_{3}^2)}{(N_{2}+N_{3})^4}},
\label{eqn:equation_for_stokes_q_u}
\end{align}
where $N_{0},\ldots,N_{3}$ are the intensities of the upper, lower, right and left spots, as shown
in Fig.~\ref{fig:optics_diagram}, and $\sigma_0,\ldots,\sigma_3$ are their uncertainties. We
estimate the uncertainty in a spot intensity $\sigma_i$ following the method outlined in
\citet{2012PASP..124..737L}:
\begin{align}
\sigma_i = & \sqrt{N_i + \sigma^2_{\rm sky}A_{\rm phot}+\frac{\sigma^2_{\rm sky}A_{\rm
phot}^2}{A_{\rm sky}}},
\end{align}
where $N_i$ is the spot intensity, $\sigma^{2}_{\rm sky} = n_{\rm sky}$ is the sky intensity
(background) in a single pixel, $A_{\rm phot}$ is the area (in pixels) of the photometry aperture,
and $A_{\rm sky}$ is the area of the background estimation annulus (see
Section~\ref{sec:PL:photometry:aperture}). The first two terms account for counting statistics of
the source and sky, while the third describes the uncertainty in the background estimation.

\begin{figure}
 \centering
 \includegraphics[width=6cm]{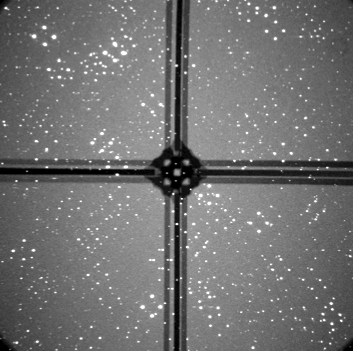}
 % example_image.png: 353x351 pixel, 72dpi, 12.45x12.38 cm, bb=0 0 353 351
\caption{An example of a RoboPol image. Each point in the sky has been mapped to four spots on the
CCD. A focal plane mask, held in place by four support legs, reduces the sky background level for
the central target. }
 \label{fig:example_image}
\end{figure}

The instrument has a large $13'\times13'$ field of view that enables relative photometry using
standard catalog sources and the rapid polarimetric mapping of compact sources in large sky areas.
While the instrument is designed to operate in the optical $V$, $R$, and $I$-bands, RoboPol
monitoring observations
are generally made using a Johnson-Cousins $R$-band filter. An example of an image from the
instrument is shown in Fig.~\ref{fig:example_image}. 

The primary scientific goal of the project is
to monitor the linear polarization of blazars, which appear as point sources at optical
wavelengths, so we optimized the instrument sensitivity for a source at
the centre of the field by using a mask in the telescope focal
plane. {\color{black} The focal plane mask has a cross-shaped aperture
  in the center where the target source is placed. The focal plane
  area immediately surrounding this aperture is blocked by the
  mask. This prevents unwanted photons from the nearby sky and sources
  from overlapping with the central target spots on the CCD,
  increasing the sensitivity of the instrument for the central
  source. The sky background level surrounding the central target spots
is reduced by a factor of 4 compared to the field sources. The focal plane
mask and its supports obscure part of the field, reducing the effective field of view.}

The polarimeter is attached to an Andor DW436 CCD camera which has an array of
$2048\times2048$ pixels and can be cooled to $-70\,$\textcelsius \, where it has 
negligible dark noise ($<0.001\,$e$^{-}$pixel$^{-1}$s$^{-1}$).

\subsubsection{Model of the instrument} \label{sec:TI:instrument:model}

Inspection of Fig.~\ref{fig:example_image} reveals that the pattern of four spots on the
CCD corresponding to a source is dependent on the location of the source in the field. This is
expected, and is due to optical distortions in the instrument. In addition to this geometric
spot-pattern effect, there are systematic errors that affect the intensity in each spot. For an
unpolarized source, the number of photons falling on each spot should be equal. However, unavoidable
imperfections in the optics result in deviations of the ratios $N_0/N_1$ and $N_2/N_3$ from 1, and
$N_0+N_1 \neq N_2+N_3$.

In our model (detailed in Appendix~\ref{sec:appendix:instrument_model}), the measured intensities
$(N_0,\ldots,N_3)$ are dependent on the location of 
the source on the CCD $(x,y)$, and are related to the true intensities $(N^*_0,\ldots,N^*_3)$ by:
\begin{align}
\nonumber N_{0} = & [1-r_{01}(x,y)]f_{01}(x)f_{P}(y)N^*_{0} \\
\nonumber N_{1} = & [1+r_{01}(x,y)]f_{01}(x)f_{P}(y)N^*_{1} \\
\nonumber N_{2} = & [1-r_{23}(x,y)]f_{23}(x)f_{P}(y)N^*_{2} \\
N_{3} = & [1+r_{23}(x,y)]f_{23}(x)f_{P}(y)N^*_{3}\label{eqn:model_for_measured_photon_counts}
\end{align}
Here $r_{01}(x,y)$ and $r_{23}(x,y)$ are functions that describe the instrumental polarization
errors -- they are the only terms that remain in the calculation of $q$ and $u$,
Eqn.~\ref{eqn:equation_for_stokes_q_u}. 
%They are well-described by second order polynomials of $x$ and $y$.
The functions $f_{01}(x)$, $f_{23}(x)$, and $f_{P}(y)$ describe the instrumental photometry
errors: the position and prism dependent optical transmission of the instrument. The form of the
error functions -- and their dependence on either $x$, $y$, or both -- were
determined by inspection of data from unpolarized standard stars. The residuals between the data
and the instrument model are uniformly distributed across the field, indicating that the model
adequately describes the spatial dependence and scale of the action of the instrument.

The model also predicts the pattern that the spots make on the CCD. As shown in
Fig.~\ref{fig:optics_diagram}, we model the distance between the spots $\Delta_x$ and $\Delta_y$,
the angle between the spots and the CCD axes $\phi_x$ and $\phi_y$, and the distance of the spots
from the intersection of their joining lines $\delta_x$ and $\delta_y$. This is modelled at every
point in the field, and is used by the pipeline to identify which spots correspond to which
astronomical source. To produce the model we take multiple exposures of an unpolarized standard
star at many locations
in the field and map the variation in the non-ideal behavior. We fit the model to the measured spot
pattern
and intensities and save the model coefficients to disk for use by the data reduction
pipeline. We re-fit the model parameters each time the instrument is removed and replaced.

\section{Pipeline} \label{sec:PL}

\subsection{Overview} \label{sec:PL:overview}

\begin{figure*}
 \centering
 \includegraphics[width=0.9\textwidth]{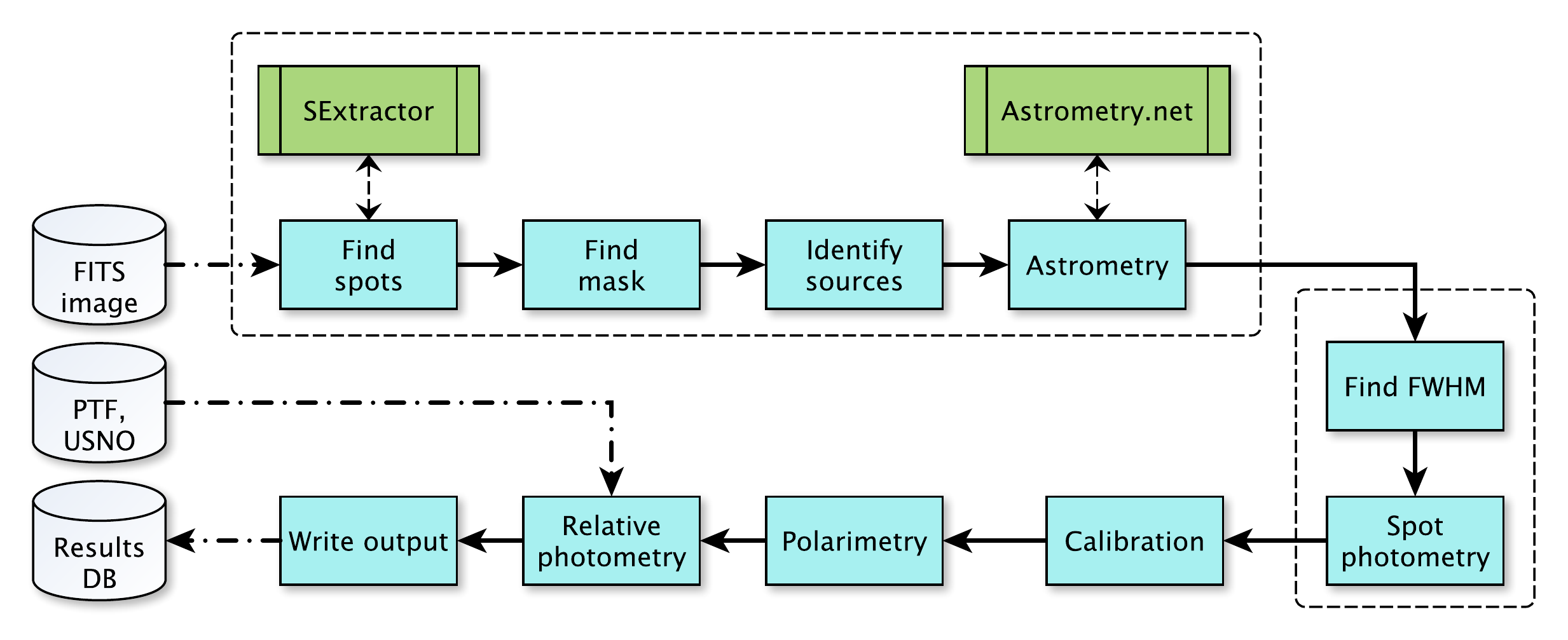}
 % pipeline_overview.pdf: 586x385 pixel, 72dpi, 20.67x13.58 cm, bb=0 0 586 385
\caption{Flow chart representation of the operation of the pipeline. The FITS image from the CCD is
first processed in a source identification step (Section~\ref{sec:PL:source_id}), in which we match
spots
to sources in the sky and calculate their astronomical coordinates. We then perform photometry on
all the identified sources (Section~\ref{sec:PL:photometry}). The measured spot counts are corrected
for
instrumental errors in the calibration step (Section~\ref{sec:PL:calibration}) before we measure the
linear polarization (Section~\ref{sec:PL:polarimetry}) and relative photometry
(Section~\ref{sec:PL:relative_photometry}).}
 \label{fig:pipeline}
\end{figure*}

The RoboPol pipeline measures the magnitude and linear polarization of every unobscured source in
the field, i.e. every source that is not obscured by the focal plane mask and its supports. 
A flow-diagram of the pipeline is shown in Fig.~\ref{fig:pipeline}.
The pipeline is written in Python, with some subroutines written in
Cython\footnote{\url{http://cython.org/}} to improve the processing
time. The operation of the pipeline can be described in five basic steps:
\begin{enumerate}
\item \textbf{Source identification}, Section~\ref{sec:PL:source_id}: Find all the spots on the CCD,
match them up to sources in the sky, solve for the world coordinate system (WCS) that describes the
image, and calculate the source coordinates from the spot coordinates.
\item \textbf{Photometry}, Section~\ref{sec:PL:photometry}: Perform aperture photometry on each of
the spots.
\item \textbf{Calibration}, Section~\ref{sec:PL:calibration}: Use the instrument model to correct
the measured spot intensities.
\item \textbf{Polarimetry}, Section~\ref{sec:PL:polarimetry}: Measure the linear polarization of
every source in the field.
\item \textbf{Relative photometry}, Section~\ref{sec:PL:relative_photometry}: Measure the $R$-band
magnitude of every source in the field by performing relative photometry using field sources.
\end{enumerate}

\subsection{Source identification} \label{sec:PL:source_id}

Every point in the sky or focal plane is mapped to four points on the CCD by the RoboPol
instrument. The first step
in the pipeline is to identify which spots on the CCD correspond to which source in the sky, i.e.
to reverse the $1\mapsto4$ mapping of the instrument.

We use SExtractor \citep{1996A&AS..117..393B} to find the
pixel coordinates of the centre of every spot on the CCD. After finding the location of the mask,
and discarding
spots whose photometry aperture is obscured by it, we find all sets of spots on the CCD that
originate from the same astronomical source (described in
Section~\ref{sec:PL:source_id:spot_matching_method} below). 

We use the central point, defined as the intersection of the line that joins the vertical spots 
and the line that joins the horizontal spots, for each set of
four spots to determine the WCS that describes the image using the \mbox{Astrometry.net}
\citep{2010AJ....139.1782L} software. We then use this WCS to transform the central pixel
coordinate for each set of four spots to a J2000
coordinate. {\color{black} The \mbox{Astrometry.net} software bases its
  astrometry on index files that are calculated from either the USNO-B
  catalog \citep{Monet:2003p2377} or the 2MASS catalog \citep{2006AJ....131.1163S}; the
  RoboPol pipeline uses the 2MASS-derived index files}\footnote{Downloaded from
    \url{http://data.astrometry.net/4200/}}. {\color{black} We have found that the
  astrometry solutions for the target source are within 3\,arcseconds
  of the catalog position 90\% of the time, independent of seeing conditions or
  position on the sky.}

\subsubsection{Spot matching method} \label{sec:PL:source_id:spot_matching_method}

A typical RoboPol exposure, such as the example shown in Fig.~\ref{fig:example_image}, contains a
large number of sources in the field. Each source forms four corresponding spots on the CCD. We
describe here a method for determining which spots on the CCD correspond to which source in the
sky, i.e. a method for finding sets of four spots automatically. We use our knowledge of
the expected spot pattern from the instrument model to do this.

Suppose we have found $M$ spots on the CCD, with pixel coordinates $(x_1,y_1),\ldots,(x_M,y_M)$. For
each spot, we can then use the instrument model (Section~\ref{sec:TI:instrument:model} and
Appendix~\ref{sec:appendix:instrument_model}) to predict
the location of the intersection of the line joining the vertical spot pair and the line joining the
horizontal spot pair, i.e. the central point. However, this requires us to know what type of spot
each spot is, i.e.
$0,1,2$, or 3. Since we do not know this a priori, we calculate where the
central point would be in each of these four cases, producing four potential central
points for each spot arranged above, below, left, and right of the spot on the CCD.

We then have a set of $4M$ predicted central points. The four spots which correspond to a 
particular source will have the same predicted central point. We search the set of predicted
central points for groups of four points that lie within a threshold of $\sim
3\,$pixels of each other, to account for centroid and model errors.

\subsection{Photometry} \label{sec:PL:photometry}

After the spots have been detected and matched to sources in the sky, we measure the intensity of 
each spot using aperture photometry. We calculate the mean FWHM across the field using 10 spots
that are bright, unblended, and unsaturated. We fit both a Gaussian and a Moffat profile
\citep{1969A&A.....3..455M}, and use the FWHM estimate from the best-fit profile.

\subsubsection{Mask detection}\label{sec:PL:mask_detection}

\begin{figure}
 \centering
 \includegraphics[width=0.47\textwidth]{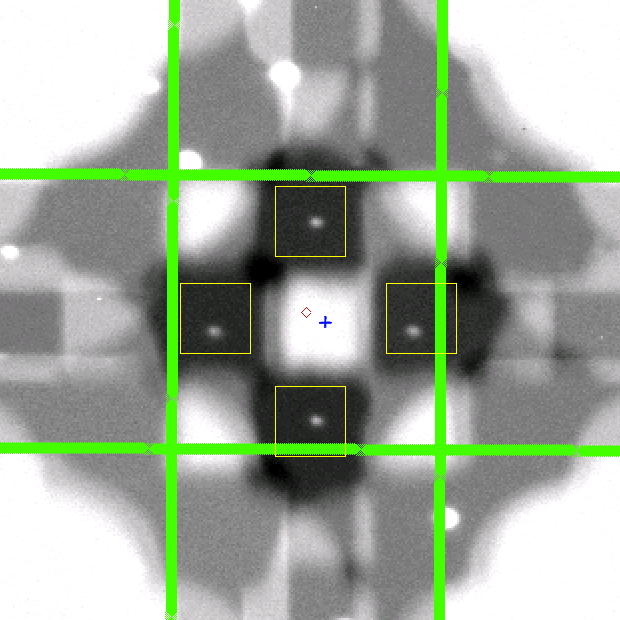}
 % center_mask.png: 620x620 pixel, 72dpi, 21.87x21.87 cm, bb=0 0 620 620
\caption{An image of the central area of the RoboPol field. The four low background areas due to the
focal plane mask, containing the four spots of the central science target, are located in the centre
of the field. The edges of the mask pattern are indicated in green. The yellow squares are the
background estimation boxes for the central science target, the red diamond is the mask centre, and
the blue cross is the pointing centre. Optical distortions result in the mask centre and pointing
centre being slightly different: the pointing centre offset from the mask centre was determined
empirically.}
 \label{fig:center_mask}
\end{figure}

We must find the exact location of the mask in order to perform aperture photometry on the central
target, and to identify and reject sources whose photometric aperture intersects with the mask. 
We find the position of the mask by fitting the known mask pattern to the image. This is done by
finding the mask pattern position that maximises the difference in background level between stripes
of pixels on either side of the mask pattern edge. Pixels contaminated by bright sources located
near the mask pattern edge are excluded from the procedure.

Knowing the geometry of the mask, we can then identify the location of
the low background areas, squares $22'' \times 22''$ in size in which the central target should be
located.
An image of the central area from a RoboPol image, with the mask pattern and low background areas
outlined, is shown in Fig.~\ref{fig:center_mask}. The mask detection is also used in the target
acquisition procedure outlined in Section~\ref{sec:CS:target_acquisition}.

\subsubsection{Aperture photometry} \label{sec:PL:photometry:aperture}

For the field sources we use circular apertures centred on each spot to measure the intensity
and an outer annulus to estimate the background level. We use the SExtractor positions of the
spot centres, and flag blended spots (these are currently not analysed by the pipeline). The focal
plane mask restricts the area around the central
target that can be used to estimate the background level, so we use a square aperture, as indicated
by the yellow boxes in Fig.~\ref{fig:center_mask}, for the central target spots to
maximize the number of pixels used in the background estimation. The location of the square
apertures is set by the location of the mask, regardless of the location of the source spots.

We estimate the background level using a method outlined in \citet{1992ASPC...23...90D}. The
background level is the mode of the smoothed distribution of all pixels that have an intensity 
within $3\sigma$ of the median level in the aperture.

\begin{figure}
 \centering
\includegraphics[width=0.47\textwidth]{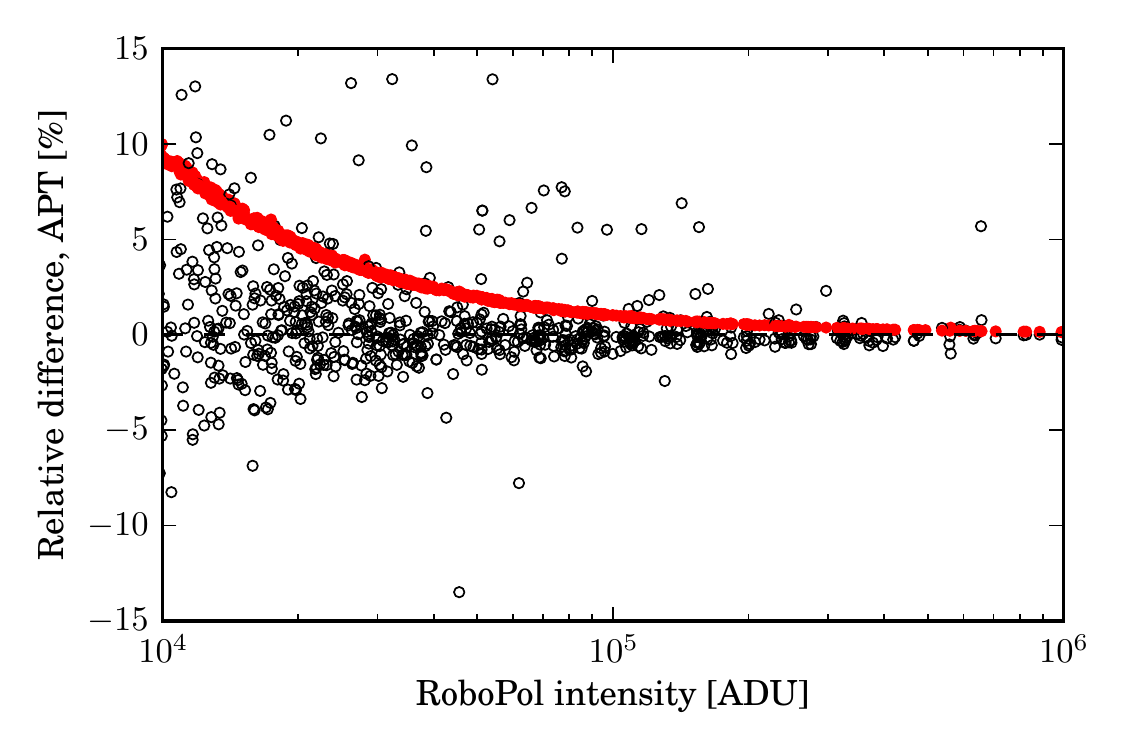}
 % robopol_intensity_vs_sextractor_and_apt_photom_method_3.pdf: 432x288 pixel, 72dpi,
%15.24x10.16cm,bb=0 0 432 288
 \caption{Comparison of the source intensity measured by the RoboPol pipeline with that measured by
Aperture Photometry Tool (APT). The APT used the same aperture settings as used by the RoboPol
pipeline. The median difference is 0.04\%. The red points are the $1\sigma$ uncertainties in the
RoboPol intensities, calculated using Poisson statistics. 
The outlier points are sources close to the focal plane mask: the RoboPol pipeline removes pixels in
the focal plane mask from the background estimation annulus, while APT does not.}
 \label{fig:photometry_result}
\end{figure}

We evaluated the performance of the RoboPol aperture photometry code by comparing its output to the
output of the Aperture
Photometry Tool (APT) software \citep{2012PASP..124..737L}. We ran APT in batch mode on a set of
RoboPol images. We fixed the photometry apertures used by APT to match those used by the RoboPol
pipeline. The results are shown in Fig.~\ref{fig:photometry_result}, where we plot the relative
difference between the RoboPol intensity and that obtained by APT. 
The APT and RoboPol results have
excellent agreement, with a median relative difference of 0.04\%. The outlier points are due to
errors in the background level estimation due to proximity of the source to the focal plane mask.

\subsection{Calibration} \label{sec:PL:calibration}

We correct the measured spot photometry for known instrumental measurement errors before we 
calculate the linear polarization and relative photometry. We use the instrument model corrections
(Section~\ref{sec:TI:instrument:model}) to correct the measured spot counts $N_{0\dots 3}$ and
obtain the corrected spot counts $N^c_{0\ldots 3}$:
\begin{align}
\nonumber N_0^c = &  \frac{N_0}{[1 - r_{01}(x_c,y_c)]  f_{01}(x_{c})f_{P}(y_{c})} \\
\nonumber N_1^c = &  \frac{N_1}{[1+r_{01}(x_c,y_c)]  f_{01}(x_{c})f_{P}(y_{c})} \\
\nonumber N_2^c = &  \frac{N_2}{[1 -  r_{23}(x_c,y_c)] f_{23}(x_{c})f_{P}(y_{c})} \\
N_3^c = &  \frac{N_3}{[1+r_{23}(x_c,y_c)] f_{23}(x_{c})f_{P}(y_{c})}
\label{eqn:corrected_spot_counts}
\end{align}
where $(x_{c},y_{c})$ is the intersection of the lines joining the vertical spot pair and the
horizontal spot pair on the CCD, the central point.

\subsection{Polarimetry} \label{sec:PL:polarimetry}

The relative linear Stokes parameters $q$ and $u$ are calculated with
Eqn.~\ref{eqn:equation_for_stokes_q_u}
using the corrected spot counts from Eqn.~\ref{eqn:corrected_spot_counts}. The linear polarization
fraction $p$ and electric vector position angle (EVPA) $\chi$ are then calculated using:
\begin{align}
p = & \sqrt{q^2 + u^2}, ~~~
\sigma_p =  \sqrt{\frac{q^2\sigma_q^2+u^2\sigma_u^2}{q^2+u^2}} \label{eqn:p}\\
\chi = & \frac{1}{2}\tan^{-1}\left(\frac{u}{q}\right), ~~~
\sigma_{\chi} =  \frac{1}{2}\sqrt{\frac{u^2\sigma_q^2+q^2\sigma_u^2}{(q^2+u^2)^2}}
\label{eqn:chi}
\end{align}
If the polarization of the source is low, i.e. $(1+q^2) \simeq (1+u^2) \simeq 1$, then the
expression for the EVPA uncertainty can be written as
\begin{align}
\sigma_{\chi} \simeq \frac{1}{2}\frac{1}{{\rm SNR}_p}, \label{eqn:EVPA_uncertainty}
\end{align}
i.e. the uncertainty in the measurement of the EVPA is
determined by the signal to noise ratio (SNR) of the polarization fraction $p$ measurement ${\rm
SNR}_{p}$.

\begin{figure}
 \centering
 \includegraphics[width=0.47\textwidth]{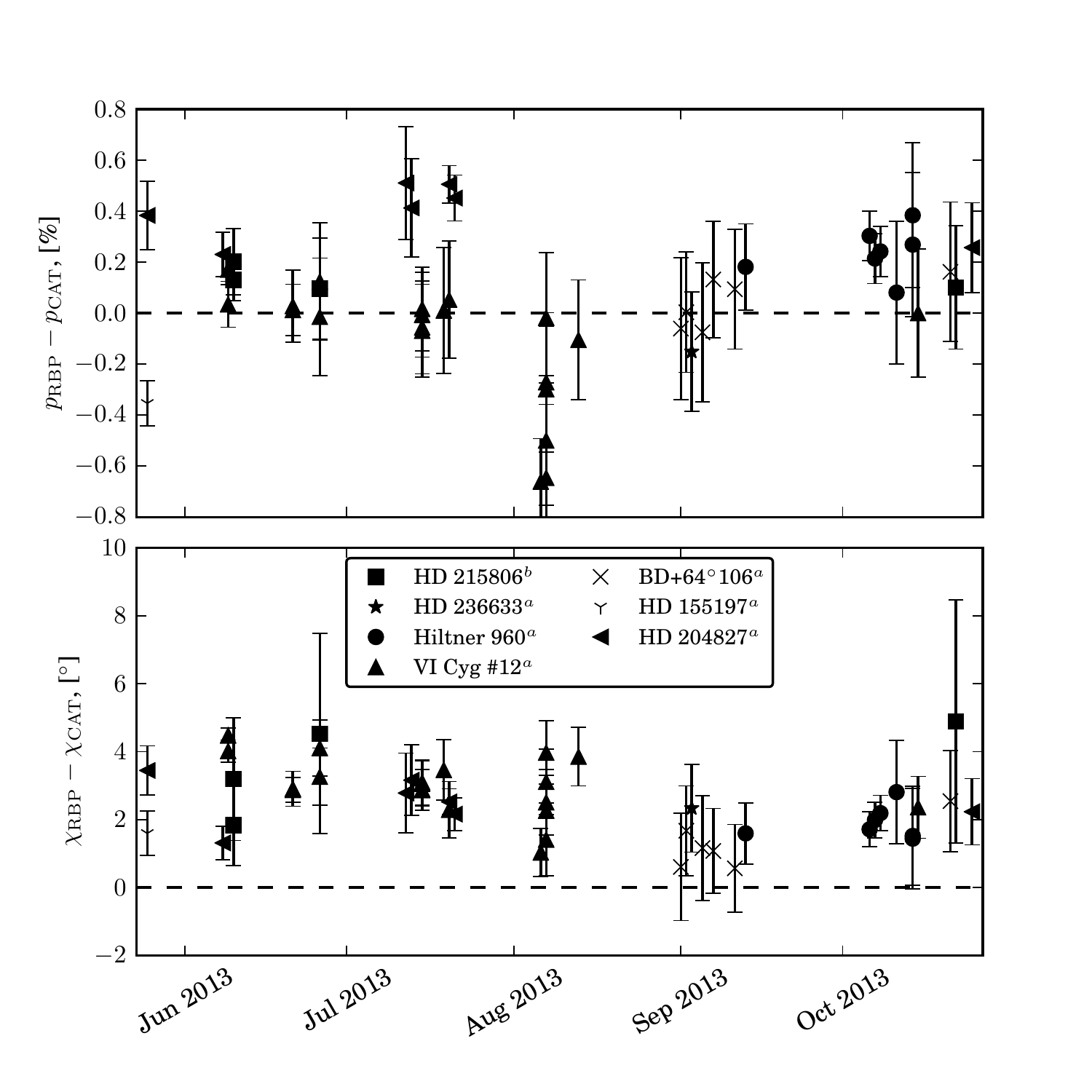}
 % calibrator_light_curves.pdf: 0x0 pixel, 0dpi, nanxnan cm, bb=
 \caption{Light curves for a selection of polarization standard stars. We plot the difference
between the RoboPol measured linear polarization percentage ($p_{\rm RBP}$) or angle ($\chi_{\rm
RBP}$), and the catalog value ($p_{\rm CAT},\chi_{\rm CAT}$). The mean values for each standard are
listed in Table~\ref{tab:polarization_measurement}, including references for the catalog values.}
 \label{fig:calibrator_light_curves}
\end{figure}

We tested the performance of the RoboPol pipeline by observing a number of polarized standard
stars with known polarization properties in the Johnson-Cousins $R$-band. The polarization standards we observed
are listed in Table~\ref{tab:polarization_measurement}. In
Fig.~\ref{fig:calibrator_light_curves} we plot light curves of  the difference between the
polarization fraction measured by the RoboPol pipeline and the catalog value, and the difference
between the RoboPol polarization angle and the catalog value. No de-biasing has been
applied, as the SNR of each measurement is large ($>$20:1). There is no systematic difference
between the RoboPol polarization percentage and the catalog value: the mean difference in
polarization percentage is $(3 \pm 5)\times 10^{-2}$. 
The polarization angles measured by the RoboPol instrument are on
average $2.31^{\circ} \pm 0.34$ larger than
the catalog angle. This is due to a rotation of the telescope polarization reference frame with
respect to the sky.

\begin{table*}
\centering
\caption{Comparison of the RoboPol pipeline results for a set of polarized
standard stars observed in the $R$-band (linear polarization percentage $p_{\rm RBP},\sigma_{\rm RBP}$ and position angle
$\chi_{\rm RBP},\sigma_{\chi,\rm RBP}$)
and their catalog values (subscript CAT). The values listed here are the unweighted means of the
measurements shown in Fig.~\ref{fig:calibrator_light_curves}.
Catalog values are from:
 $^{a}$, \citet{1992AJ....104.1563S}, 
 $^b$, \citet{1992ApJ...386..562W}. }
\label{tab:polarization_measurement}
\begin{tabular}{|c|c|c|c|c|c|c|c|c|}
\hline
%Source & $p_{\rm RBP}$ [\%] & $\sigma_{\rm RBP}$ & $p_{\rm CAT}$ [\%] & $\sigma_{\rm CAT}$ 
Source & $p_{\rm RBP}$ [\%] & $\sigma_{p,{\rm RBP}}$ & $\chi_{\rm RBP}$ [$^{\circ}$] &
$\sigma_{\chi,{\rm RBP}}$ & $p_{\rm CAT}$ [\%] & $\sigma_{p,{\rm CAT}}$ & $\chi_{\rm CAT}$
[$^{\circ}$] & $\sigma_{\chi,{\rm CAT}}$ 
\\ \hline
VI Cyg \#12$^{a}$ & 7.78 & 0.05 & 119.2 &  0.2 & 7.893 & 0.037 & 116.23 & 0.14 \\
HD 236633$^{a}$ & 5.22 & 0.23 & 95.4 &  1.3 & 5.376 & 0.028 & 93.04 & 0.15 \\
Hiltner 960$^{a}$ & 5.45 & 0.08 & 56.4 &  0.4 & 5.210 & 0.029 & 54.54 & 0.16 \\
BD+64$^{\circ}$106$^{a}$ & 5.19 & 0.10 & 98.0 &  0.6 & 5.150 & 0.098 & 96.74 & 0.54 \\
HD 204827$^{a}$ & 5.29 & 0.06 & 61.6 &  0.3 & 4.893 & 0.029 & 59.10 & 0.17 \\
HD 155197$^{a}$ & 3.92 & 0.09 & 104.5 &  0.7 & 4.274 & 0.027 & 102.88 & 0.18 \\
HD 215806$^{b}$ & 1.96 & 0.09 & 69.6 &  1.3 & 1.830 & 0.040 & 66.00 & 1.00 \\ \hline
\end{tabular}
\end{table*}

\subsection{Relative photometry} \label{sec:PL:relative_photometry}

We measure the brightness of the objects in the RoboPol field relative to a set of
non-variable reference sources
in the field. This requires a reliable reference photometric catalog. There are two
catalogs that have significant overlap with our sources, the PTF (Palomar Transient Factory) 
$R$-band catalogue \citep{2012PASP..124..854O} and the USNO-B1.0 catalog \citep{Monet:2003p2377}.
However, we have found the USNO-B1.0 magnitudes to be unsuitable for use as photometric standards
due to their marginal photometric quality.

The PTF $R$-band catalogue magnitudes are of a high quality, with very low systematic errors of
$\sim0.02\,$mag, but the data were taken using a Mould $R$ filter and the resultant catalog
magnitudes are in the PTF photometric system \citep{Ofek:2012cj}. 
{\color{black} We transform the $R_{\rm PTF}$ magnitude to the
  Johnson-Cousins system using the transformation provided in
  \citet{Ofek:2012cj}, Equation~6. Since we do not know the color of
  each object in the field a-priori, we use the median color term
  $\alpha_{c,R} = 0.214$ for all sources in the PTF catalog to obtain the transformation: }
\begin{align}
R_{\rm PTF} \simeq R_{c} +
0.086\times\left(R_{c}-I_{c} \right) + 0.124 . 
\end{align}
We then used SDSS data\footnote{\url{http://cas.sdss.org/astro/en/tools/crossid/upload.asp}} to study the
colors of the PTF reference objects. We found 13,091 PTF reference sources with corresponding SDSS
$r-i$ colors, with the mean of the color distribution being 0.18 and the width (standard deviation)
0.19. We therefore ignore the negligible color-dependent part of the transformation and use the
relationship:
\begin{align}
R_{c} \simeq R_{\rm PTF} - 0.124. \label{eqn:PTF_mag_correction}
\end{align}
{\color{black} This approximate photometric transformation will
  become redundant when we complete our catalog of Johnson-Cousins reference
  magnitudes, as discussed in Section~\ref{sec:conclusion}.}

We identify all reference sources in the RoboPol frame that are uncontaminated, i.e. that are not
blended sources and that do not have any sources in their background estimation annulus. We find
their catalog magnitude and convert it to a flux using the zero point for the Johnson-Cousins
photometric system. We find the best-fit line to the total source intensity (sum of the four spot
intensities $\sum_{i=1}^{4}N_{i}^{c}$) vs flux for the reference sources, and use this relationship
to convert the total intensity for all the sources in the frame to an $R$-band magnitude. We measure
the standard deviation of the difference between the RoboPol magnitude and the catalog magnitude
for the reference sources, and call this the ``standards'' uncertainty. This systematic
uncertainty is the same for every source in the field. The uncertainty in the magnitude of a source
is then the quadrature sum of the statistical uncertainty for that source (SNR in intensity
measurement) and the ``standards'' uncertainty.

\begin{figure}
 \centering
 \includegraphics[width=0.47\textwidth]{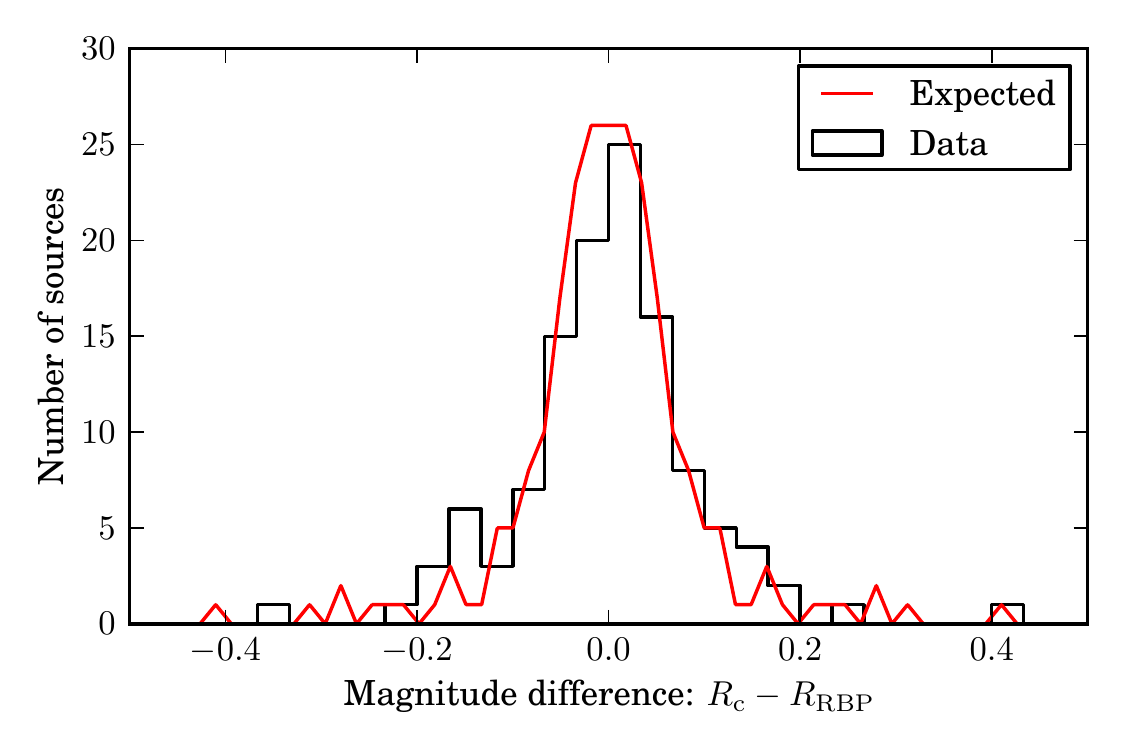}
 % magnitude_difference_distribution.pdf: 324x216 pixel, 72dpi, 11.43x7.62 cm, bb=0 0 324 216
 \caption{Distribution of the difference between the RoboPol-measured $R$-band magnitude $R_{\rm
RBP}$ and the PTF $R$-band magnitude (corrected using Eqn.~\ref{eqn:PTF_mag_correction}) $R_{\rm
c}$. The red curve shows the distribution of the expected magnitude uncertainty calculated as
$R_{\rm RBP}\dfrac{\sigma_N}{N}$ (where $N$ is the source intensity and $\sigma_N$ is its
uncertainty), mirrored around 0. The difference in the magnitudes is consistent with the level
expected from photon counting statistics; no systematic difference is evident. }
 \label{fig:magnitude_difference_distribution}
\end{figure}

In Fig.~\ref{fig:magnitude_difference_distribution} we show
the distribution of the difference between the RoboPol $R$-band magnitude and the PTF catalog
$R$-band magnitude (corrected using Eqn.~\ref{eqn:PTF_mag_correction}) for a set of RoboPol field
sources. The magnitudes are very similar, and the scatter in the difference is consistent with that
expected from the SNR in the spot photometry measurement.

\section{Control system} \label{sec:CS}

\subsection{Overview} \label{sec:CS:overview}

\begin{figure*}
 \centering
 \includegraphics[width=\textwidth]{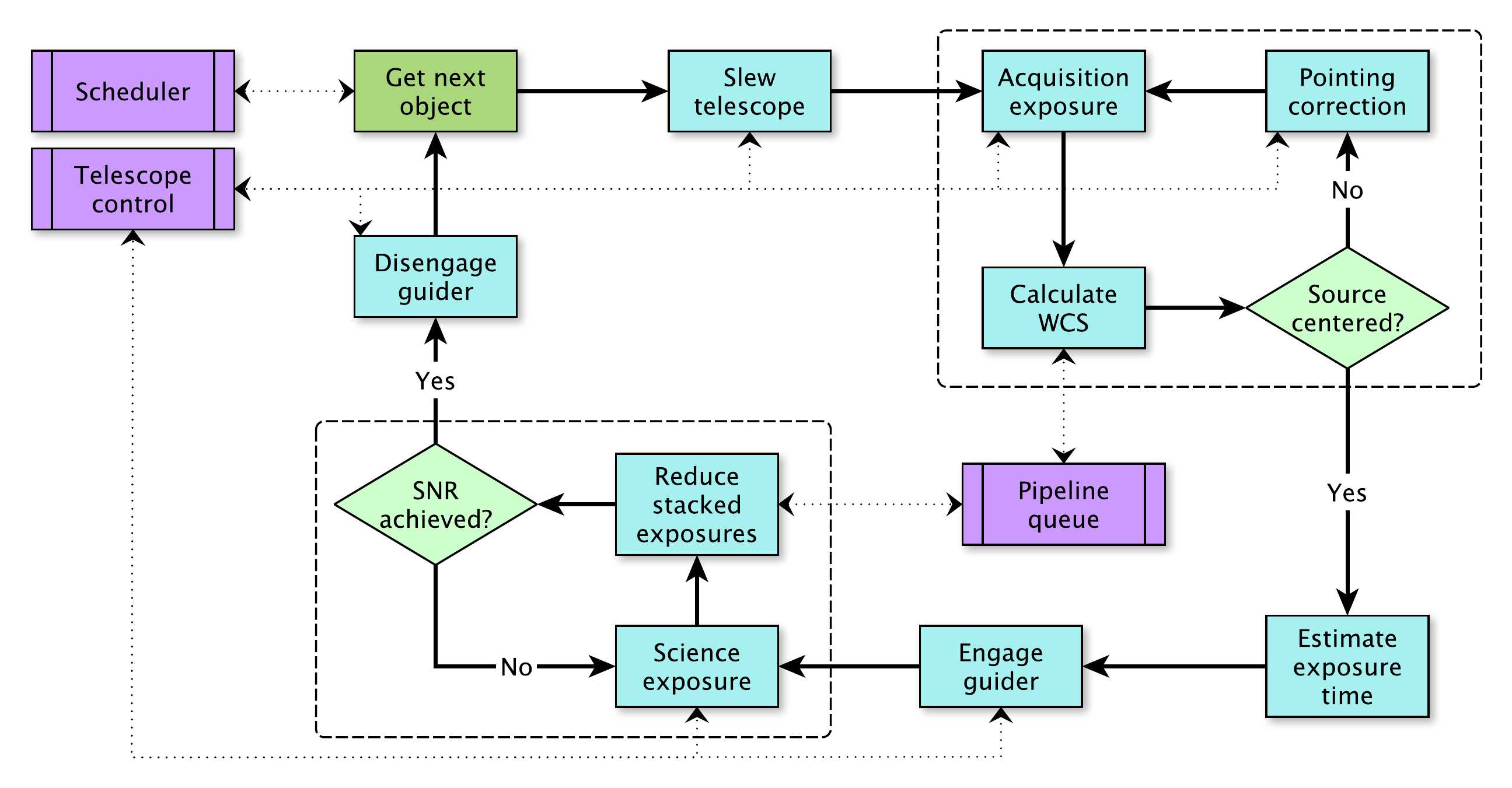}
 % control_system_overview.pdf: 746x389 pixel, 72dpi, 26.32x13.72 cm, bb=0 0 746 389
\caption{Flow chart representation of the main observing loop. The observing loop gets the next
object from the scheduler process (Section~\ref{sec:CS:scheduling}) and instructs the telescope
control
process to slew to the target. A target acquisition loop (Section~\ref{sec:CS:target_acquisition})
then
ensures that the science target is centred in the mask. The source is then observed until the SNR
goal is reached (Section~\ref{sec:CS:exposure}). The loop then acquires the next object from the
observing queue. }
 \label{fig:control_system_overview}
\end{figure*}

The RoboPol control system is designed with high observing efficiency and dynamic scheduling as
prime goals. High efficiency is
achieved by full automation of the observing process, and dynamic adjustment of the exposure time
for a target to reach a specified SNR goal.

The control system operates the Skinakas $1.3$-m telescope robotically during RoboPol observing
sessions, and allows full manual control of the telescope the rest of the
time. As described in Section~\ref{sec:TI:telescope}, the control of the telescope subsystems is
spread over several computers running a variety of operating systems. The RoboPol control system is
written in Python and consists of a number of independent processes running on these computers,
communicating with each other over ethernet using TCP sockets.

A simplified flow chart of the main observing loop in the master control process is shown in
Fig.~\ref{fig:control_system_overview}. Some of the other independent processes are shown in purple.
The control system processes are:
\begin{description}
\item[\textbf{Master}:] Control the observing process.
\item[\textbf{Scheduler}:] Provide the next object to observe (Section~\ref{sec:CS:scheduling}).
\item[\textbf{Pipeline queue}:] Analyse the FITS images from the instrument and provide the
science target magnitude and linear polarization to the master and scheduler processes.
\item[\textbf{Gamma-ray data pipeline} (not shown):] Process the gamma-ray data
provided by the \emph{Fermi} LAT telescope offline and provide the latest data to the scheduler
process.
\item[\textbf{Telescope control}:] Interface with the mount, dome, and focus control through
the TCS computer, control of the RoboPol filter wheel and CCD.
\item[\textbf{GUI} (not shown):] A graphical interface to the control system to provide the
telescope operator with feedback and allow manual intervention if necessary.
\item[\textbf{Weather} (not shown):] Monitor a weather station to provide information to the
watchdog processes and for logging.
\item[\textbf{Watchdogs} (not shown):] Monitor and maintain the stability of the control system.
\end{description}

In addition to the fully-automated main observing loop, the control system runs an automated focus
routine (Section~\ref{sec:CS:autofocus}) several times during the night, automatically acquires
flat-field exposures (Section~\ref{sec:CS:autoflat}) to monitor dust contamination of the optics,
and
has a target-of-opportunity mode that can interrupt the main observing loop to observe, for
instance, gamma-ray burst optical afterglows.

All exposures made by the control system are stored on disk at the telescope and transferred once a
day to
servers at the University of Crete. From there the data are distributed over the internet to the
partner institutions for redundant backup. A database of light curves for all the sources in every
RoboPol field is maintained at the University of Crete.

\subsection{Dynamic scheduling} \label{sec:CS:scheduling}

The RoboPol control system is designed to allow dynamic scheduling. At the start of each night the
scheduler process
produces a nominal schedule of the sources from the RoboPol catalog that are due to be observed. As
each source is observed its measured magnitude and linear polarization are passed to the scheduler
process to allow changes to the schedule to be made, if necessary. This dynamic response mode is not
being used in the first observing season while we gather the data necessary to characterize the
behavior of our sources and develop the algorithms to reliably identify interesting behavior.
Details of the dynamical scheduler will be reported in future papers.

\subsection{Target acquisition} \label{sec:CS:target_acquisition}

The pointing requirements for the RoboPol instrument are very stringent: we require the science
target to be within $2''$ of the pointing centre of the mask. We cannot achieve this precision
with a blind slew to a source, so the control system contains a target acquisition loop to centre
the source in the mask before taking the science exposures.

\begin{figure}
 \centering
 \includegraphics[width=0.47\textwidth]{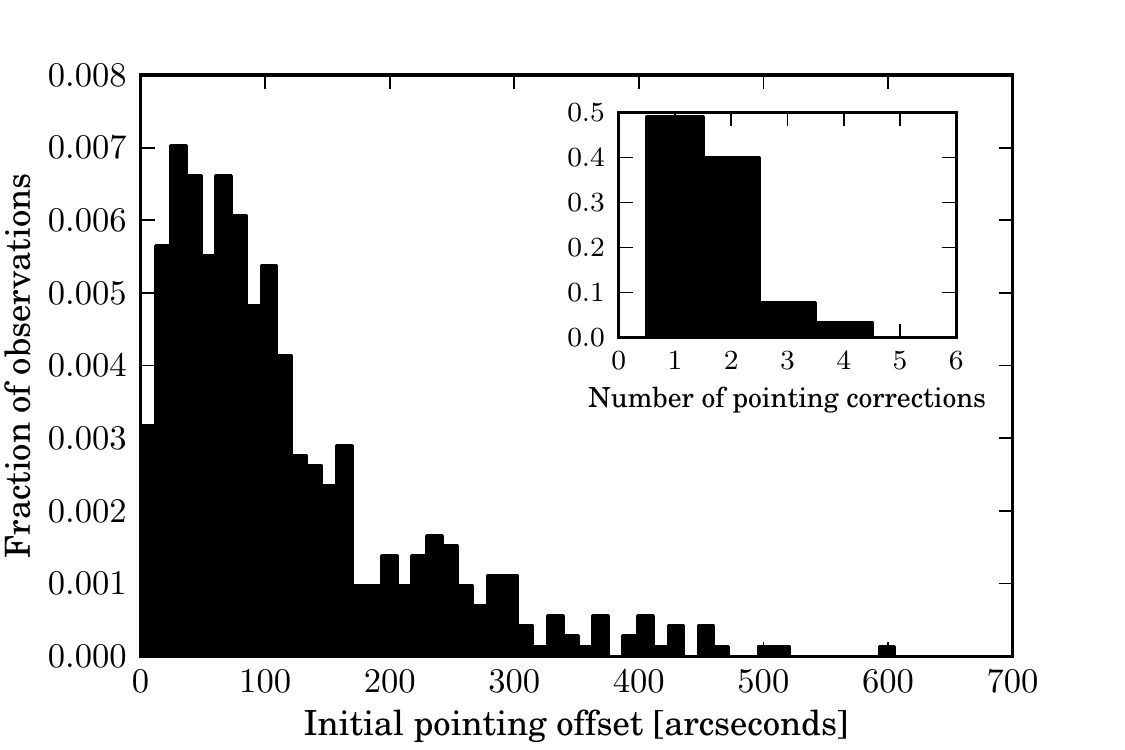}
 % target_acquisition_performance.pdf: 324x216 pixel, 72dpi, 11.43x7.62 cm, bb=0 0 324 216
\caption{The initial pointing offset of the field centre from the commanded position. (inset) About
90\% of sources require two or fewer pointing corrections to be properly centred.}
 \label{fig:target_acquisition_performance}
\end{figure}

After the initial telescope slew, the control system takes a short exposure of the field. This is
processed using the pipeline to find the mask location and to calculate the WCS that describes the
frame. A pointing correction that would place the target coordinates at the mask pointing centre is
calculated. This correction is applied and another short exposure is taken. This loop is repeated
until the target source is properly located. The performance of the target acquisition system is
shown in Fig.~\ref{fig:target_acquisition_performance}. Most initial slews are within $\sim2'$ of
the commanded position, and $\sim90$\% of sources require two or fewer pointing corrections to be
properly placed in the mask.

It is not necessary for the central target source to be visible in a single exposure for this
procedure to work. As long as there are enough stars in the field for the pipeline to calculate the
WCS for the frame, the location of the source in the field can be calculated and the appropriate
pointing correction applied.

\subsection{Dynamic exposure time} \label{sec:CS:exposure}

Both the polarization and magnitude of blazars are highly variable at optical wavelengths. For
greater observing efficiency we expose only long enough to reach a target SNR of 10:1 in $p$,
which equates to an uncertainty in the EVPA
of $\sim 2.86^{\circ}$ (see Eqn.~\ref{eqn:EVPA_uncertainty}). Because the blazar emission can change
significantly
from night
to night (and even within a night), we calculate the necessary exposure time to reach our SNR goal
from the data as we gather it.

We use the final target acquisition exposure to provide an initial guess for the required exposure
time. We calculate the amount of time needed to collect 250,000~photons in total from the source,
which we have found gives an SNR in $p$ of $\sim$10:1 in a $\sim 3$\% polarized source under average
observing conditions at Skinakas. We
then take a number of science exposures; as the science exposures are accumulated we run the
pipeline on the stacked image and update the estimate of the required observing time. We stop
observing once the SNR goal is reached, or when the total exposure time has reached 40~minutes.

\subsection{Autofocus} \label{sec:CS:autofocus}

The RoboPol instrument is optimized to measure the linear polarization of point sources. The
control system contains an autofocus mode that takes a series of exposures at different focus
positions. It then finds the focus position that produces the lowest median FWHM across the field.
While the FWHM does  vary across the field, the minimum in the median FWHM corresponds to the same
focal position as the minimum in the FWHM of the central target, and the curve of median FWHM vs
focus position has lower noise than the curve for a single source. 
This procedure is run at the beginning and mid-way through each night.

\subsection{Autoflats} \label{sec:CS:autoflat}

The control system automatically takes flat-field exposures at dawn or dusk, which are used to
track the presence of dust in the telescope optics and its effect on the performance of the
instrument.

We select an observing location for the flat-field exposures by requiring that the distance of the
target flat-field sky area from the Moon be more than
$50^{\circ}$ and that the distance from the horizon be more than $40^{\circ}$, thereby limiting the
gradient of the background to $<1$\% across our field \citep{1996PASP..108..944C}. The control
system selects as the target sky area the point on the line of declination $\delta=+32^{\circ}$ that
meets these criteria and has the greatest summed distance from the Moon and the horizon.

According to \citet{1993AJ....105.1206T} the logarithm of the brightness of the sky changes linearly
with time, with possible deviations due to atmospheric dust. We have found that the sky
brightness light curve is better described by a $2^{\rm nd}$ order polynomial. We take a series of
short exposures of the sky every $120\,$s to characterize the median sky brightness light curve. 
Once the changing sky brightness is adequately characterized, we calculate the optimum time to
start taking the flat-field exposures such that we get a median background count of $\sim
10000\,$ADU per pixel ($\sim 1/3$ of the non-linear point for this CCD) in the first flat-field exposure. We then take a series of $3-10\,$s
exposures while varying the pointing location of the telescope, which is used to calculate the master
flat-field image.

\section{Conclusions} \label{sec:conclusion}

We have described the data reduction pipeline and control system developed for the RoboPol project.
We have shown that the aperture photometry using circular apertures performed by the RoboPol
pipeline produces results that are indistinguishable from those obtained using the standard aperture
photometry tool APT \citep{2012PASP..124..737L}. Our aperture photometry code has a
substantially faster processing time than APT, though it should be noted that APT was not
designed with processing speed as a primary goal. By using our own code we are also able to use a
square background
estimation aperture for the central target, thereby taking full advantage of the focal plane mask.

Most optical polarimeters use a rotating polarization element to remove instrumental effects
from the polarization measurement. The RoboPol instrument does not; we instead take a single
exposure and use a model of the instrumental effects (Appendix~\ref{sec:appendix:instrument_model})
to correct the measured spot intensities
before calculating the source polarization. The instrument model is derived from observations of
unpolarized standard stars at multiple locations in the RoboPol field of view. We used the RoboPol
pipeline to analyse observations of a set of polarized standard stars and found that the measured
polarizations matched the catalog polarizations to within the statistical error.

The measurement of the magnitude of the sources in a RoboPol image is obtained by relative
photometry against photometric standards in the field. The only catalogue of photometric
standards of sufficient quality and sky coverage to be suitable for use with the RoboPol data is the
PTF $R$-band catalogue \citep{2012PASP..124..854O}. However, it does not include the fields around
every source in the RoboPol sample {\color{black} and is not in the same photometric
system as the RoboPol data}, so we are in the process of taking the necessary data to extend
the PTF $R$-band catalogue to cover all RoboPol sources {\color{black}
  in the Johnson-Cousins photometric system}. All the RoboPol data will be reprocessed
using our new catalogue of standards once it is complete. We have shown that the magnitudes measured
by the RoboPol pipeline are consistent with those in the PTF catalogue.

The RoboPol control system is written to allow dynamic scheduling. It analyses each image as
it is taken and sends the results to the scheduler. A primary goal of the RoboPol project
is to use this information to respond immediately to important changes in a source's behavior
without human intervention. However, knowing what changes in a source's emission are important
requires us to first characterize their behavior. In the first observing season we are taking the
data necessary to perform this characterization, and will report on the resulting design of the
dynamical scheduler in future papers. Due to the modular design of the RoboPol control system we
will be able to change the scheduling code with minimal effort.

\section*{Acknowledgments}

The RoboPol project is a collaboration between Caltech in the USA,
MPIfR in Germany, Toru\'{n} Centre for Astronomy in Poland, the University of
Crete/FORTH in Greece, and IUCAA in India.
The U. of Crete group acknowledges support by the ``RoboPol'' project, which is implemented under
the ``Aristeia'' Action of the  ``Operational Programme Education and Lifelong Learning'' and is
co-funded by the European Social Fund (ESF) and Greek National Resources, and by the European
Comission Seventh Framework Programme (FP7) through grants PCIG10-GA-2011-304001 ``JetPop'' and
PIRSES-GA-2012-31578 ``EuroCal''.
This research was supported in part by NASA grant NNX11A043G and NSF grant AST-1109911, and by the
Polish National Science Centre, grant number 2011/01/B/ST9/04618.
K.\,T. acknowledges support by the European Commission Seventh Framework Programme (FP7) through
the Marie Curie Career Integration Grant PCIG-GA-2011-293531 ``SFOnset''.
M.\,B. acknowledges support from the International Fulbright Science and Technology Award.
T.\,H. was supported in part by the Academy of Finland project number
267324.
I.M. was supported in this research through a stipend from the
International  Max Planck Research School (IMPRS) for Astronomy and
Astrophysics at the Universities of Bonn  and Cologne.
This research made use of Astropy, \url{http://www.astropy.org}, a community-developed core Python
package for Astronomy \citep{astropy}.

\bibliographystyle{mn2e}
\bibliography{bibliography_manual,bibliography_export}

%\clearpage
%\newpage

\appendix
\section{Instrument model} \label{sec:appendix:instrument_model}

\begin{figure}
 \centering
 \includegraphics[width=7cm]{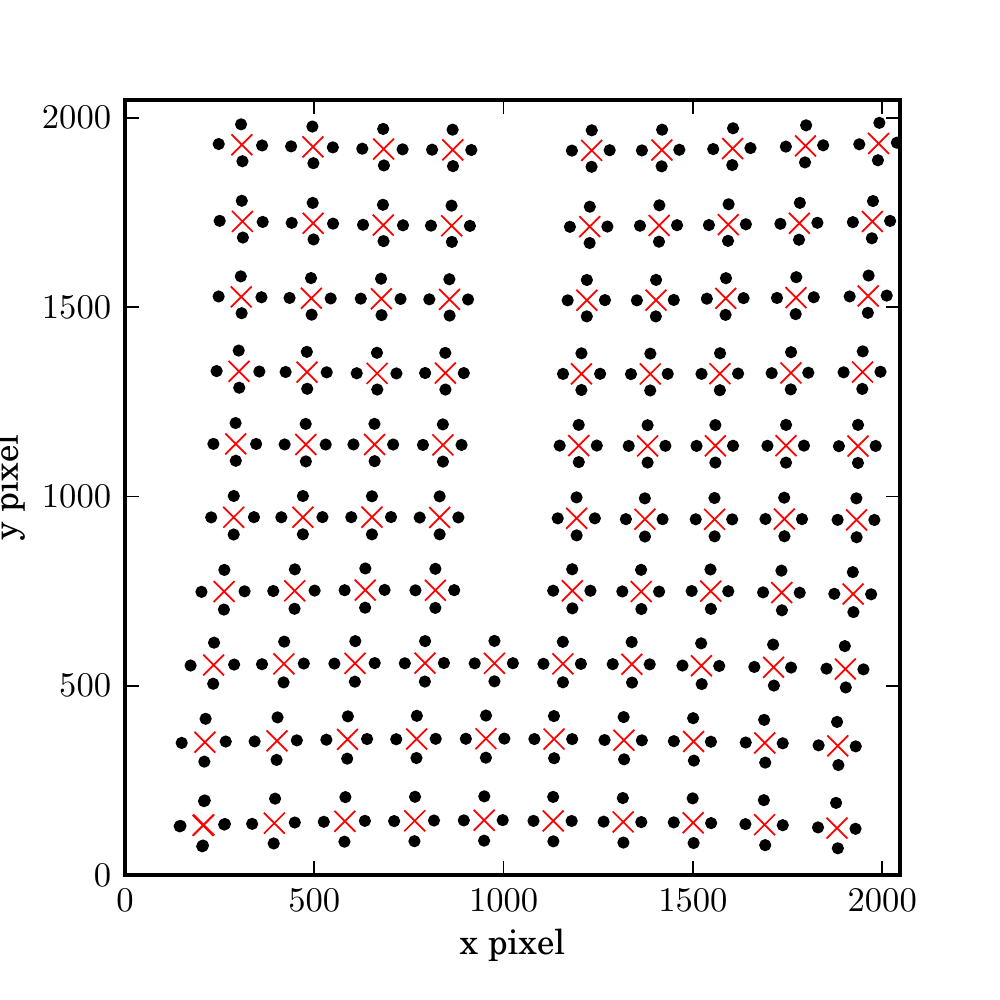}
 % CAL_star_raster_location_map.pdf: 288x288 pixel, 72dpi, 10.16x10.16 cm, bb=0 0 288 288
 \caption{A plot showing the location of the standard unpolarized star HD~154892 in a series of
exposures. The individual spots are indicated by black dots, and the central point by a red
cross.}
 \label{fig:raster_map_plot}
\end{figure}

The instrument model describes two separate behaviors of the RoboPol receiver. These are the
variation in the spatial pattern made by the spots on the CCD, and the effect on the intensity of
each spot. The data used to generate the model come from a series of exposures of a standard
unpolarized star. In each exposure, the telescope pointing is stepped by $1\,$arcminute, thereby
sampling a grid of points in the field of view with the standard source.
Fig.~\ref{fig:raster_map_plot} shows the locations of the standard star in a series of such
exposures. 

Since the intrinsic magnitude and polarization of the source does not change over the
course of the exposures, any changes in the observed magnitude or polarization of the source are due
to aberrations in the combined telescope and instrument optics. We can then model the corrections
to the spot intensities that will result in a source of zero polarization and constant magnitude
regardless of where in the field of view it is located.

The model described here is agnostic about the source of the aberrations that it corrects for.
It is purely empirical: it corrects for the observed behavior with as few parameters as necessary,
regardless of the physical source of the aberrant behavior. The functional forms used in the model
were selected by best-fit to the data, rather than derived from a physical model of the optics.

\subsection{Spatial model}

The spatial model predicts the location of the four spots on the CCD, given the location of the
source $(x,y)$. As shown in Fig.~\ref{fig:optics_diagram}, this pattern is described by 6
numbers. The distance between the horizontal spots is given by $\Delta_x(x,y)$, and between the
vertical spots it is $\Delta_y(x,y)$. The distance from the right-spot to the central
point $(x,y)$ is given by $\delta_x(x,y)$, and from the upper-spot it is
$\delta_y(x,y)$. Finally, the angle between the CCD $x$-axis and the horizontal line is
$\phi_{x}(x,y)$ and the angle between the CCD $y$-axis and the vertical line is
$\phi_{y}(x,y)$.

\begin{figure*}
\centering
\includegraphics[width=\textwidth]{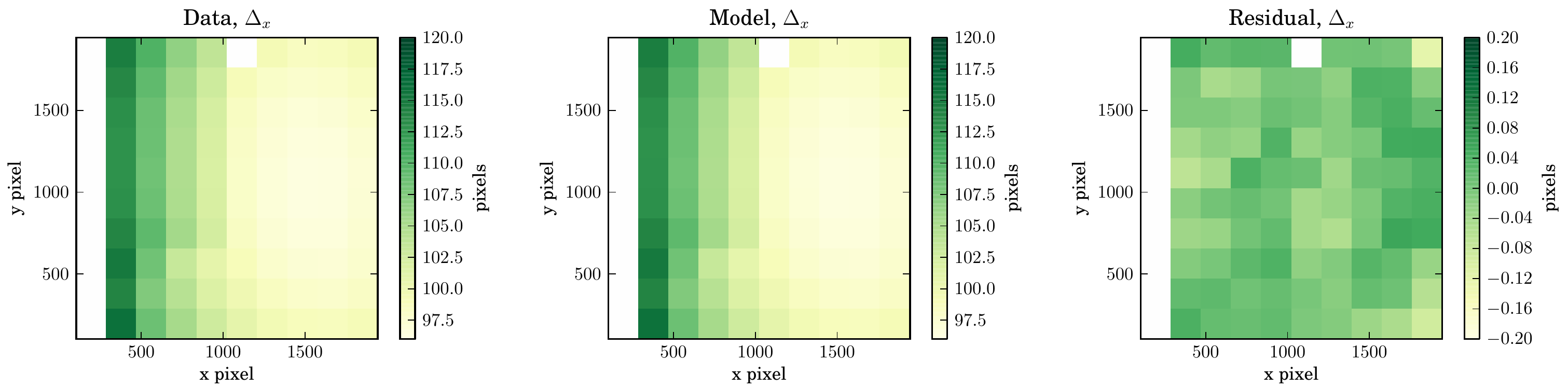}
\caption{The data (left), best-fit model (centre), and residuals (right) for the quantity $\Delta_x$
in the instrument spatial pattern model. Note the change in color scale for the residual plot.
Areas in the plot with no data, due to imperfect coverage of the field as shown in
Fig.~\ref{fig:raster_map_plot}, are blank.}
 \label{fig:spatial_model:delx}
\end{figure*}

\begin{figure*}
\includegraphics[width=\textwidth]{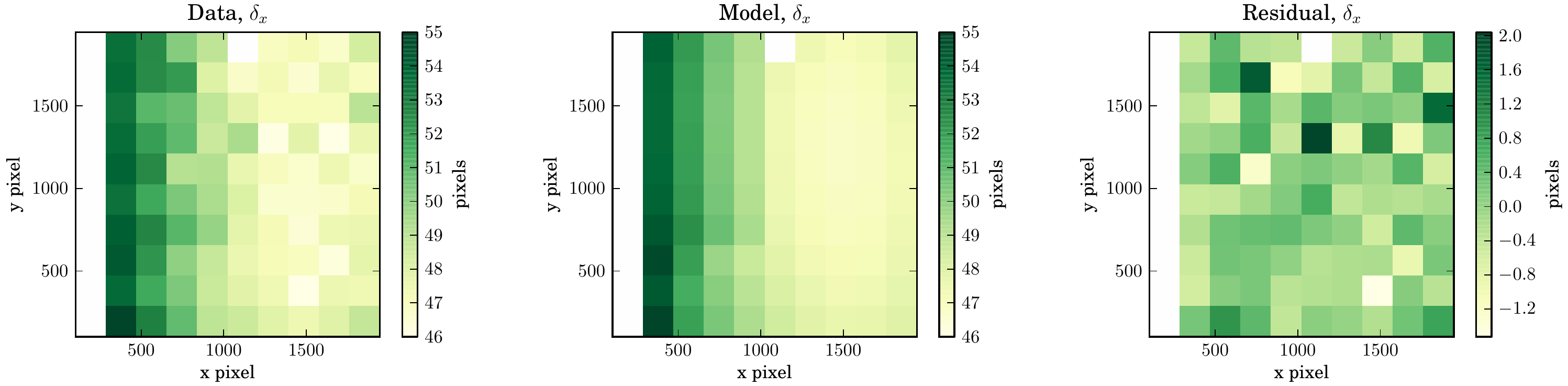}
\caption{The data (left), best-fit model (centre), and residuals (right) for the quantity $\delta_x$
in the instrument spatial pattern model. Note the change in color scale for the residual plot.}
 \label{fig:spatial_model:dx}
\end{figure*}

\begin{figure*}
\includegraphics[width=\textwidth]{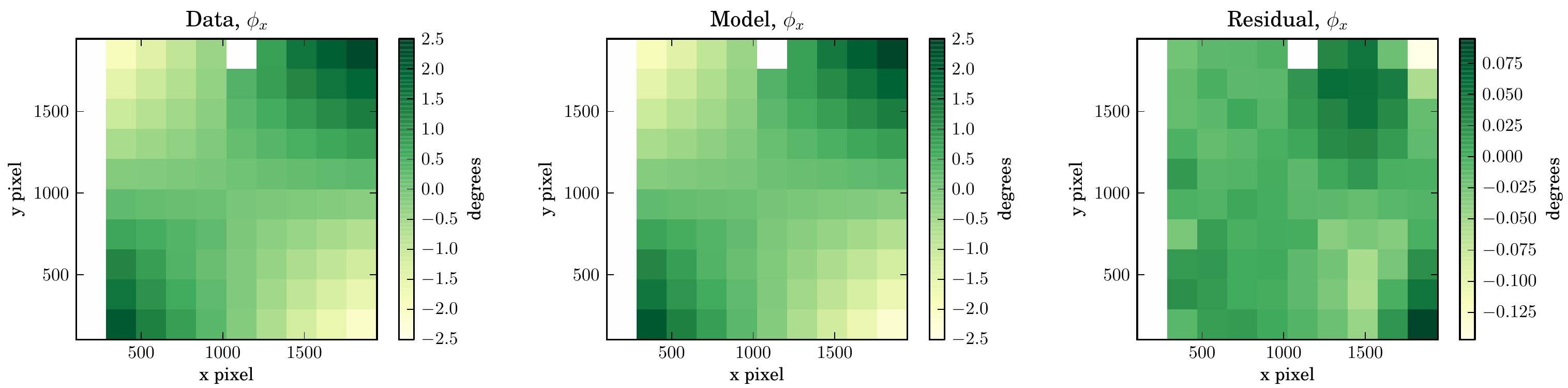}
\caption{The data (left), best-fit model (centre), and residuals (right) for the quantity $\phi_x$
in the instrument spatial pattern model. Note the change in color scale for the residual plot.}
 \label{fig:spatial_model:phix}
\end{figure*}

The left panel in Figs.~\ref{fig:spatial_model:delx}, \ref{fig:spatial_model:dx}, and
\ref{fig:spatial_model:phix} shows the measurement of these quantities for the $x$ subscript. The
$y$ version looks similar. In these plots we have split the CCD in to 100 cells and have plotted
the average quantity in each cell. We have found empirically that the data are well-fitted by these
functional forms:
\begin{align}
\Delta_x(x,y) = & P^{3}_{\Delta x,1}(x_{c}) + P^{2}_{\Delta x,2}(y_{c}) \\
\Delta_y(x,y) = & P^{3}_{\Delta y,1}(y_{c}) + P^{2}_{\Delta y,2}(x_{c}) \\
\delta_x(x,y) = & P^2_{\delta x,1}(x) + P^2_{\delta x,2}(y) \\ 
\delta_y(x,y) = & P^2_{\delta y,1}(y) + P^2_{\delta y,2}(x) \\
\phi_x(x,y) = & a_0 [ (x-a_1) + (x-a_1)(y-a_2) +  (y-a_2) ] + a_3 \label{eqn:model:phi_x} \\
\phi_y(x,y) = & b_0 [ (x-b_1) + (x-b_1)(y-b_2) +(y-b_2) ] + b_3.\label{eqn:model:phi_y}
\end{align}
Here, $P^{N}(x)$ is a polynomial of order $N$ in $x$, and $a_i,b_i$ are coefficients for the $\phi$
expressions.

The best-fit model for each case are shown in the centre panel of
Figs.~\ref{fig:spatial_model:delx},
\ref{fig:spatial_model:dx}, and \ref{fig:spatial_model:phix}, and the residual (difference between
the data and the model in each cell) is shown in panel (c). The fits are generally excellent, with
the residuals for the $\delta$ parameters being noise-dominated. Some coherent structure remains in
the residuals for the $\Delta$ and $\phi$ quantities, but the level of the residuals are low enough
that the spot matching method described in Section~\ref{sec:PL:source_id:spot_matching_method}
works, and there is hence no need to add additional detail to the model.

\subsection{Intensity model}

\begin{figure*}
 \centering
\includegraphics[width=0.7\textwidth]{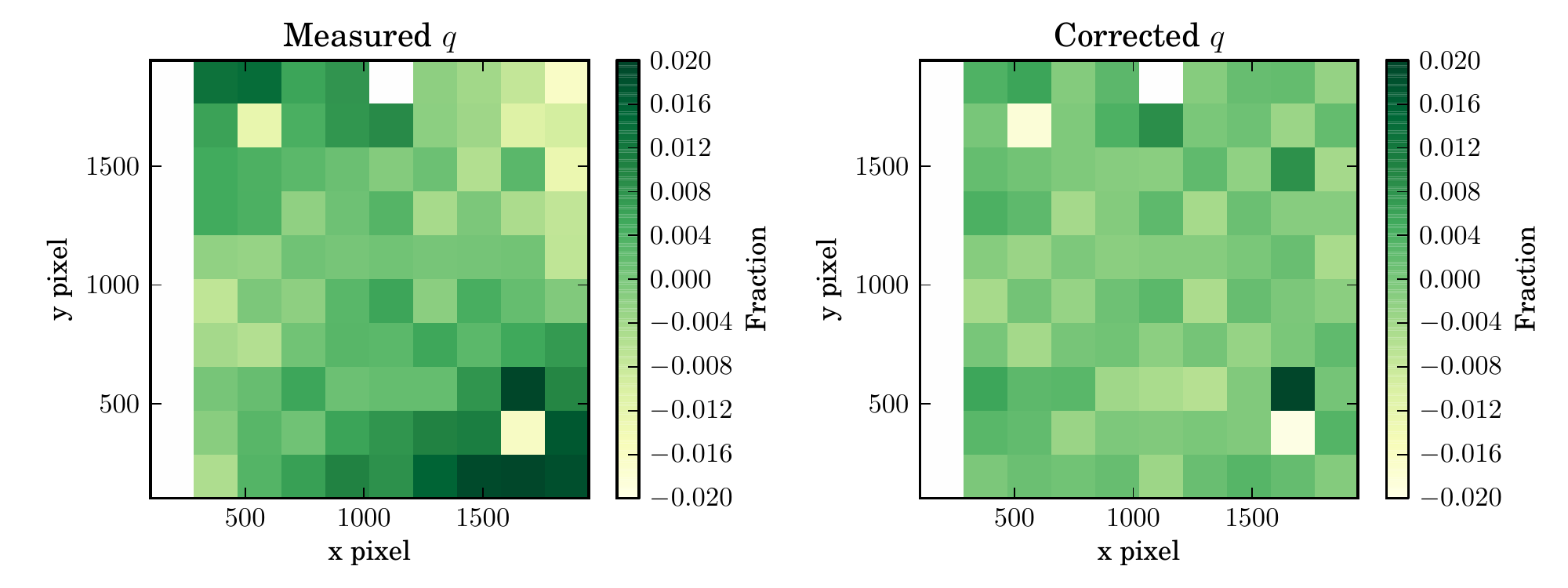}
\caption{The uncorrected (left) and corrected (right) relative Stokes $q$ parameter, which
correspond to before and after applying the instrument intensity model to the data, respectively.}
 \label{fig:intensity_model:q}
\end{figure*}

\begin{figure*}
 \centering
\includegraphics[width=0.7\textwidth]{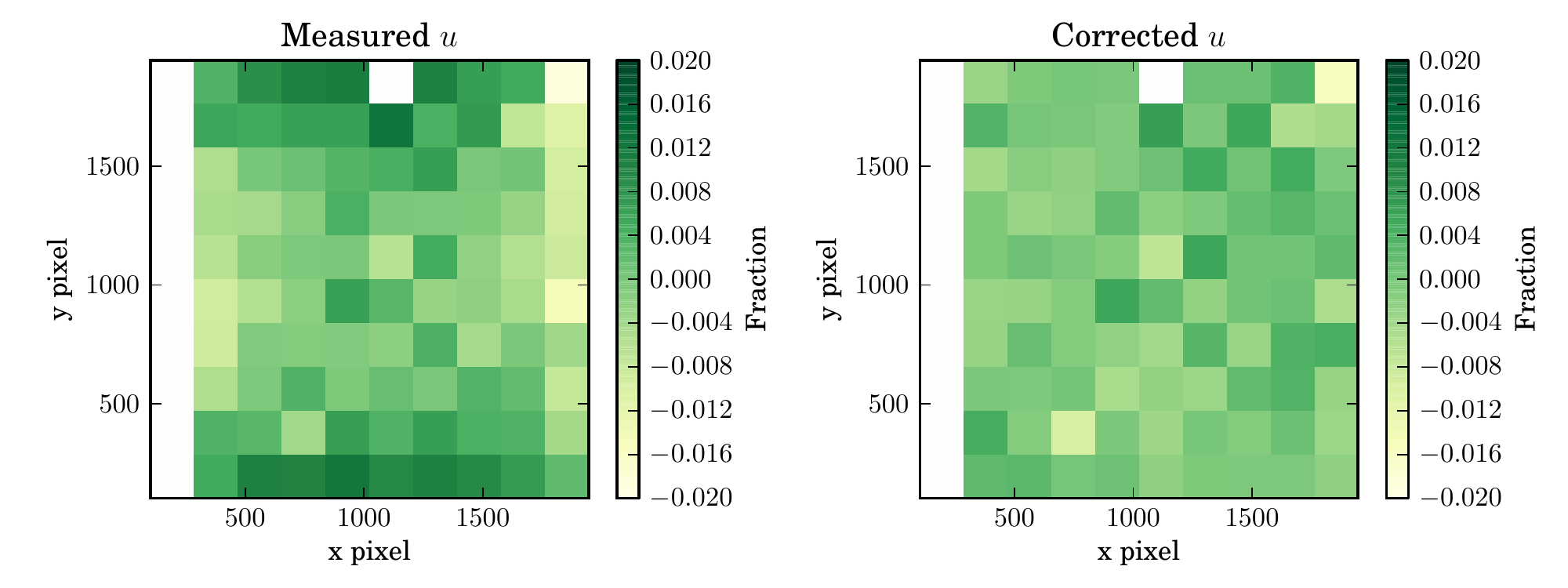}
\caption{The uncorrected (left) and corrected (right) relative Stokes $u$ parameter, which
correspond to before and after applying the instrument intensity model to the data, respectively.}
 \label{fig:intensity_model:u}
\end{figure*}

The instrument intensity model 
is used to correct the measured spot intensities for systematic errors that affect the polarimetry
and relative photometry measurements. We model the measured spot intensities as:
\begin{align}
N_{0} = & [1-r_{01}(x,y)]f_{01}(x)f_{P}(y)N_{0}^* \\
N_{1} = & [1+r_{01}(x,y)]f_{01}(x)f_{P}(y)N_{1}^* \\
N_{2} = & [1-r_{23}(x,y)]f_{23}(x)f_{P}(y)N_{2}^* \\
N_{3} = & [1+r_{23}(x,y)]f_{23}(x)f_{P}(y)N_{3}^*,
\end{align}
where $N_{i}$ is the measured spot intensity and $N_{i}^*$ is the true spot intensity for spot
$i=0,\ldots,3$. The instrumental polarization is determined by the parameters $r_{01}(x,y)$ and
$r_{23}(x,y)$. These parameters describe the ratios of the intensities in spots 0/1 and spots 2/3
respectively. As in the spatial instrument model, we determined the best-fit functional forms
empirically:
\begin{align}
r_{01}(x,y) = & 1 + R_{01,1}(x,y) + R_{01,2}(x,y) \\
\nonumber & {\rm where:} \\
\nonumber R_{01,1}(x,y) = & a_0[ (x-a_1) + (x-a_1)(y-a_2) +\\
 &  (y-a_2)] + a_3 \\
R_{01,2}(x,y) = & b_0[(y - b_1)^2 - (x - b_2)^2].\label{eqn:model:r01}
\end{align}
The same functional forms are used to determine the model for $r_{23}(x,y)$.
Figs.~\ref{fig:intensity_model:q} and \ref{fig:intensity_model:u} show the measured and corrected
relative Stokes parameters. Large position-dependent systematic errors are evident in the
uncorrected plots, while the corrected plots have the expected mean of 0 with no systematic errors.
The coefficients $a_i,b_i$ have no relation to those in Equations~\ref{eqn:model:phi_x} and
\ref{eqn:model:phi_y}.

\begin{figure*}
\includegraphics[width=0.7\textwidth]{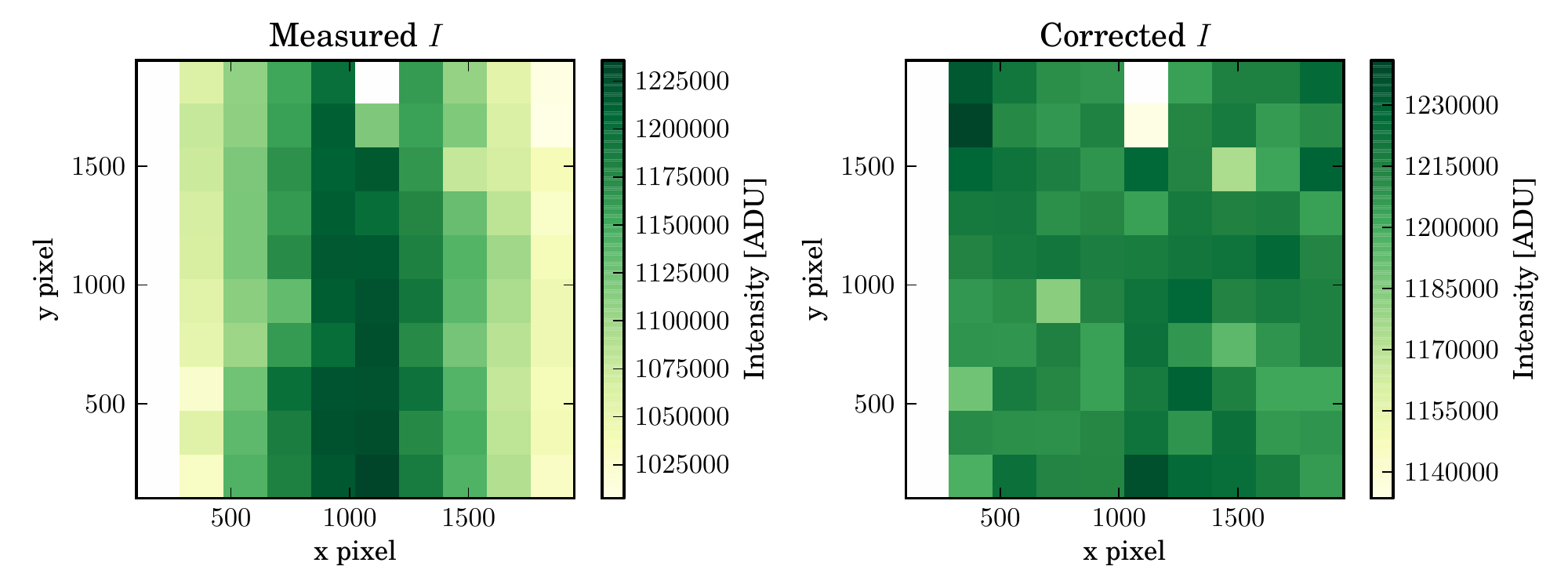}
 \caption{The uncorrected (left) and corrected (right) total source intensity, which correspond
 to before and after applying the instrument intensity model to the data, respectively. Note the
change in color scale for the corrected plot.}
 \label{fig:intensity_model:phot}
\end{figure*}

The functions $f_{01}(x)$, $f_{23}(x)$, and $f_{P}(y)$ describe the instrumental photometry
errors: the position and prism dependent optical transmission of the instrument. The functional
form that describes $f_{01}(x)$ is:
\begin{align}
f_{01}(x) = & \left\{
     \begin{array}{ll}
       y_1 & : x \ge x_{\rm cr}\\
       y_2 & : x < x_{\rm cr}
     \end{array}
   \right. \\
\nonumber & {\rm where:} \\
y_1 = &\frac{a_0}{\pi}\left[2\arccos f - \sin\left(2\arccos f\right)\right] \\
y_2 = &  hx^2 + gx + k \\
\nonumber & :f =  \frac{a_1(x-x_{\rm cr})}{2048} \\
\nonumber & :g = \frac{(a_3 - 1)a_0}{(a_2/2 + x_{\rm cr}^2/(2a_2) - x_{\rm cr})} \\
\nonumber & :h = \frac{-g}{2a_2} \\
\nonumber & :k = a_0 - h x_{\rm cr}^2 - g x_{\rm cr} 
\end{align}
$y_1$ captures the effect of partly blocking an aperture stop. $y_2$ fits the response in the region
where the aperture
stop is not blocked with a $2^{\rm nd}$ order polynomial. The parameter $x_{\rm cr}$ is the point on
the CCD where we transition from a blocked to an unblocked aperture stop. A similar function
describes $f_{23}(x)$. The function $f_{P}(y)$ captures a dependence on $y$ that affects all four
spot intensities equally, and is described by a second-order polynomial. The model coefficients
$a_i$ are not related to those used in Equations~\ref{eqn:model:phi_x}, \ref{eqn:model:phi_y}, and
\ref{eqn:model:r01}.
Fig.~\ref{fig:intensity_model:phot} shows the uncorrected and corrected total source intensity
(sum of all four spot intensities). The corrected source intensity is free of systematic error.

%\clearpage

\label{lastpage}

\end{document}